\documentclass[longauth, traditabstract]{aa} 
\usepackage{graphicx}
\usepackage{txfonts}
\usepackage{natbib}
\begin{document}

\title{{\em Herschel} observations of EXtra-Ordinary Sources: the present and future of spectral 
surveys with {\em Herschel}/HIFI 
\thanks{{\em Herschel} is an ESA space
    observatory with science instruments provided by European-led
    Principal Investigator consortia and with important participation
    from NASA.}}

\author{
E.~A.~Bergin,\inst{1}
T.~G.~Phillips,\inst{2}
C.~Comito,\inst{3}
N.~R.~Crockett,\inst{1}
D.~C.~Lis,\inst{2}
P.~Schilke,\inst{3,4}
S.~Wang,\inst{1}
T.~A.~Bell,\inst{2}
G.A.~Blake,\inst{5}
B.~Bumble,\inst{6}
E.~Caux,\inst{7,8}
S.~Cabrit,\inst{9}
C.~Ceccarelli,\inst{10}
J.~Cernicharo,\inst{11}
F.~Daniel,\inst{11,12}
Th.~de~Graauw,\inst{13}
M.-L.~Dubernet,\inst{14,15}
M.~Emprechtinger,\inst{2}
P.~Encrenaz,\inst{12}
E.~Falgarone,\inst{12}
M.~Gerin,\inst{12}
T.~F.~Giesen,\inst{4}
J.~R.~Goicoechea,\inst{11}
P.~F.~Goldsmith,\inst{6}
H.~Gupta,\inst{6}
P.~Hartogh,\inst{16}
F.~P.~Helmich,\inst{13}
E.~Herbst,\inst{17}
C.~Joblin,\inst{7,8}
D.~Johnstone,\inst{18}
J.~H.~Kawamura,\inst{6}
W.~D.  Langer,\inst{6} 
W.~B. Latter,\inst{19}
S.~D.~Lord,\inst{19}
S.~Maret,\inst{10}
P.~G.~Martin,\inst{20}
G.~J.~Melnick,\inst{21}
K.~M.~Menten,\inst{3}
P.~Morris,\inst{19}
H.~S.~P. M\"uller,\inst{4}
J.~A.~Murphy,\inst{22} 
D.~A.~Neufeld,\inst{23}
V.~Ossenkopf,\inst{4,13}
L..~Pagani,\inst{9}
J.~C.~Pearson,\inst{6}
M.~P\'erault,\inst{12}
R.~Plume,\inst{24}
P.~Roelfsema,\inst{13}
S.-L.~Qin,\inst{4}
M.~Salez,\inst{9}
S.~Schlemmer,\inst{4}
J.~Stutzki,\inst{4}
A.~G.~G.~M.~Tielens,\inst{25}
N.~Trappe,\inst{22}
F.~F.~S.~van der Tak,\inst{13}
C.~Vastel,\inst{7,8}
H.~W.~Yorke,\inst{6}
S.~Yu,\inst{6}
\and
J.~Zmuidzinas\inst{2}
}
\institute{Department of Astronomy, University of Michigan, 500 Church Street, Ann Arbor, MI 48109, USA \\
\email{ebergin@umich.edu}
\and California Institute of Technology, Cahill Center for Astronomy and Astrophysics 301-17, Pasadena, CA 91125 USA
\and Max-Planck-Institut f\"ur Radioastronomie, Auf dem H\"ugel 69, 53121 Bonn, Germany 
\and I. Physikalisches Institut, Universit\"at zu K\"oln,
              Z\"ulpicher Str. 77, 50937 K\"oln, Germany
\and California Institute of Technology, Division of Geological and Planetary Sciences, MS 150-21, Pasadena, CA 91125, USA
\and Jet Propulsion Laboratory,  Caltech, Pasadena, CA 91109, USA
\and Centre d'\'etude Spatiale des Rayonnements, Universit\'e de Toulouse [UPS], 31062 Toulouse Cedex 9, France
\and CNRS/INSU, UMR 5187, 9 avenue du Colonel Roche, 31028 Toulouse Cedex 4, France
\and  LERMA \& UMR8112 du CNRS, Observatoire de
 Paris, 61, Av. de l'Observatoire, 75014 Paris, France
\and Laboratoire d'Astrophysique de l'Observatoire de Grenoble, 
BP 53, 38041 Grenoble, Cedex 9, France.
\and Centro de Astrobiolog\'ia (CSIC/INTA), Laboratiorio de Astrof\'isica Molecular, Ctra. de Torrej\'on a Ajalvir, km 4
28850, Torrej\'on de Ardoz, Madrid, Spain
\and LERMA, CNRS UMR8112, Observatoire de Paris and \'Ecole Normale Sup\'erieure, 24 Rue Lhomond, 75231 Paris Cedex 05, France
\and SRON Netherlands Institute for Space Research, PO Box 800, 9700 AV, Groningen, The Netherlands
\and LPMAA, UMR7092, Universit\'e Pierre et Marie Curie,  Paris, France
\and  LUTH, UMR8102, Observatoire de Paris, Meudon, France
\and MPI f\"ur Sonnensystemforschung, D 37191 Katlenburg-Lindau,
Germany
\and Departments of Physics, Astronomy and Chemistry, Ohio State University, Columbus, OH 43210, USA
\and National Research Council Canada, Herzberg Institute of Astrophysics, 5071 West Saanich Road, Victoria, BC V9E 2E7, Canada 
\and Infrared Processing and Analysis Center, California Institute of Technology, MS 100-22, Pasadena, CA 91125
\and Canadian Institute for Theoretical Astrophysics, University of Toronto, 60 St George St, Toronto, ON M5S 3H8, Canada
\and Harvard-Smithsonian Center for Astrophysics, 60 Garden Street, Cambridge MA 02138, USA
\and  National University of Ireland Maynooth. Ireland
\and  Department of Physics and Astronomy, Johns Hopkins University, 3400 North Charles Street, Baltimore, MD 21218, USA
\and Department of Physics and Astronomy, University of Calgary, 2500
University Drive NW, Calgary, AB T2N 1N4, Canada
\and
Leiden Observatory, Leiden University, PO Box 9513, 2300 RA Leiden, The Netherlands\label{inst2}
}


\abstract{We present initial results from the {\em Herschel} GT key program: {\em Herschel 
observations of EXtra-Ordinary Sources} (HEXOS) and outline the promise 
and potential of spectral surveys with {\em Herschel}/HIFI.  The HIFI instrument offers 
unprecedented sensitivity, as well as continuous spectral coverage across the gaps 
imposed by the atmosphere, opening up a largely unexplored wavelength regime to high-resolution spectroscopy.    We show the spectrum of Orion KL between 480 and 560 GHz
and from 1.06 to 1.115 THz.   From these data, we confirm that HIFI separately measures the dust continuum and
spectrally resolves emission lines in Orion KL.   Based on this capability we demonstrate that
the line contribution to the broad-band continuum in this molecule-rich source is $\sim$20-40\% below 1 THz and declines to a few percent
at higher frequencies.    We also tentatively identify multiple transitions of  HD$^{18}$O in the spectra.  The first detection of this rare isotopologue in the interstellar medium suggests that HDO emission is optically thick in the Orion hot core with HDO/H$_2$O $\sim$ 0.02.   We
discuss the implications of this detection for the water D/H ratio in hot cores.
}

   \titlerunning{Spectral Surveys with {\em Herschel}}
	\authorrunning{Bergin et al.}
   \maketitle
%

\section{Introduction}

Massive star-forming regions are characterized by a rich molecular emission
spectrum \citep[e.g.][]{hvd09}.   One profitable method of 
exploring gas physics and chemistry has been to survey the spectrum within the mm/sub-mm  atmospheric windows.    
This allows for an unbiased look at the chemical composition and, via the multitude of detected lines, for
inferring physical parameters such as the temperature and density.   Of particular note in this regard are 
the Orion and Sgr B2 star-forming regions. 
The hot cores (Orion KL, Sgr B2 N $+$ M) within these two clouds have been the 
subject of intense scrutiny with numerous spectral surveys
revealing a spectrum dominated by organic molecules \citep{blake87, nummelin00, comito_850, persson07}.  For a complete reference list, see \citet{tcpg10} and \citet{belloche09}.  Such observations have illustrated the rich chemical complexity that is attributed to grain surface reactions, which is revealed as the newly formed star heats the dust and releases frozen ices \citep{ehren_araa, hvd09}.

Surveying large regions of spectrum with near uniform sensitivity comes
at a cost in telescope time owing to overheads induced by limited bandwidth,
receiver tuning,  calibration errors induced by
atmospheric variations, relative calibration accuracy between bands, and changes in the pointing.  
For the next several years, this will change with the opportunity provided by the 
{\em Herschel}/HIFI instrument.   HIFI has a built in mode to rapidly scan large portions 
of the spectrum with the high sensitivity provided
by a space-based platform.   In addition, HIFI opens a large portion of the sub-mm/far-IR spectrum ($158 - 610$ $\mu$m) 
for high-resolution ($\lambda/\Delta\lambda > 10^6$) spectroscopy.   This is crucial for uncharted spectral
territory, but also for detecting the lines of H$_2$O, a key molecular constituent.


We present here some initial results from the  HIFI spectral scans of Orion KL obtained as part 
of the {\em Herschel} guaranteed time key program,  {\em Herschel observations
of EXtra-Ordinary Sources} (HEXOS).    The aim of this paper is not only to outline the goals and  sample results 
from one particular {\em Herschel} key program, 
but also to highlight the tremendous utility of {\em Herschel}/HIFI for rapidly obtaining high-resolution spectra in a rich region
of the electromagnetic spectrum.   In \S 2 we discuss  HEXOS goals and methodology.
In \S 3 we show some of the first spectral scans, demonstrate the ability of HIFI to resolve the dust continuum, and 
 comment on the observed line-to-continuum ratio in the far infrared.  In \S 4 we present one of the unexpected results from an unbiased spectral view in the detection of HD$^{18}$O in the Orion KL hot core.

\vspace{-0.5cm}

\section{HIFI Spectral Surveys and HEXOS}

The observations discussed here are part of the HEXOS guaranteed-time key program on the
{\em Herschel} satellite.  The HEXOS observational program consists primarily of complete HIFI spectral scans of Orion KL, Sgr B2 (N), Sgr B2 (M), Orion S, and the Orion Bar.   This is supplemented by a number of deep integrations and small maps.   The broad goals of the HEXOS program are to (1) define the submillimeter spectrum of dense warm molecular gas, (2)  provide a near complete chemical assay and cooling census of star-forming gas in a variety of environments, (3) explore the physical perspective offered by observations of hundreds of lines of a single molecule,
(4) use the high excitation lines to probe the chemical and physical state of gas in close proximity to the newly formed massive star(s), and (5) search the spectrum for new molecular constituents and potentially identify the bending transitions of polycyclic aromatic hydrocarbons.

It is important to note that the stable thermal space-based environment and fast tuning of {\em Herschel}/HIFI is crucial for
achieving these goals.      The specifics of the HIFI instrument have been discussed by \citet{degraauw10} and the {\em Herschel} Space Observatory by \citet{pilbratt10}.  
The principle adopted to plan a spectral scan for HEXOS is to achieve
a uniform coverage in source-intrinsic terms.  That is, we  adopted a
source intrinsic desired rms and  a source size (determined by
interferometric observations) and then calculated the goal rms for
HIFI taking the source coupling to the main beam of the telescope into
account.  Likewise, we estimated all sensitivities with a fixed
resolution in velocity (1 km s$^{-1}$ in this instance).  Based on this the spectral scan of Orion KL from $\sim 480$ GHz to 1900 GHz (with some gaps) took $\sim$ 45 hours (including two separate pointings in bands 6 and 7).   For reference, to cover 40 GHz with comparable spectral resolution at the Caltech Submillimeter Observatory (CSO) to 30 mK rms required $\sim$ 36 hrs (Widicus-Weaver, priv. comm.).  The CSO data has higher spatial resolution (at comparable frequencies), but this comparison illustrates that HIFI samples a vast region of the spectrum at comparable sensitivity with a substantial reduction in telescope time.

  A full HIFI spectral scan ultimately covers $>$ 1000 GHz with  $\sim$1 MHz spectral resolution obtained by the same instrument, minimizing relative calibration uncertainties.  This enables a direct comparison of lines spanning a wide range of frequencies and opens the capability for exploring an extensive range of topics, such as those addressed by the selection of papers in this  A\&A issue.
  

\section{Observations}

The HIFI observations were obtained in March and April 2010
using the dual beam-switch (DBS) mode pointed towards the Orion Hot
Core at $\alpha_{J2000} = 5^h35^m14.3^s$ and $\delta_{J2000} =
-5^{\circ}22'36.7''$.   We used the normal chop setting for the SiS bands and fast chop
for the HEB bands with reference beams approximately
3$^{\prime}$ east and west.  The
Wide Band Spectrometer provides a spectral resolution of 1.1 MHz
 over a 4~GHz IF bandwidth.   The data were reduced using HIPE pipeline version 2.4 and are calibrated to T$_A^*$ scale.  The velocity calibration of HIFI data is good to $\sim 0.5$ km s$^{-1}$.

\begin{figure}
\resizebox{\hsize}{!}{\includegraphics{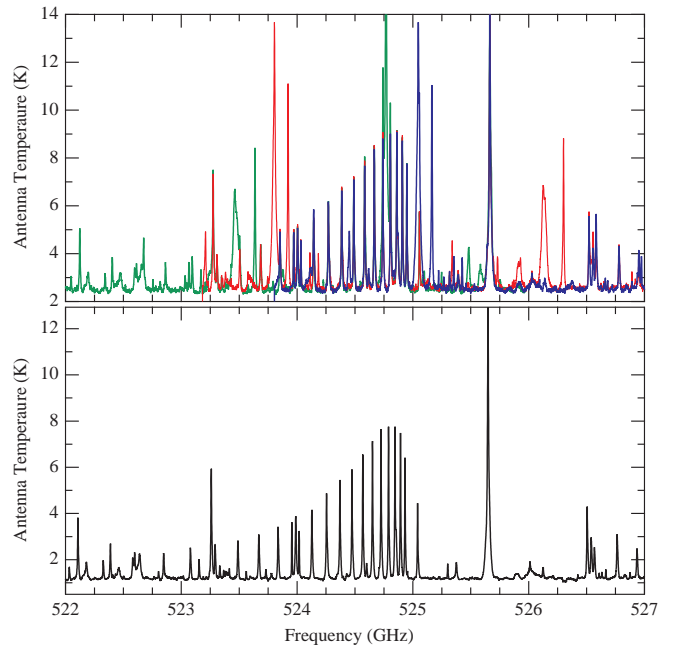}}
\caption{Illustration of the double sideband deconvolution within a section of the band 1a scan obtained with a redundancy of 4.  {\em Top:} Three different LO settings shown with separate colors. Lines in the other sideband move at the different LO settings.  {\em Bottom:} Single sideband spectrum after the deconvolution.  The obvious band in the spectrum is a Q-branch of methanol ($J_{-4,0}-J_{-3,0}$E; $J =$ 10 - 25), which is discussed by \citet[][]{wang10}.}
\label{fig:dsb}
\end{figure}

HIFI operates as a double sideband system where, in the conversion to frequencies detectable by the spectrometers, spectral features in the opposite sideband appear superposed at a single frequency. 
As part of the spectral scan observation, different settings of the local oscillator (LO) are observed and the double sideband is deconvolved to isolate the observed sideband \citep{cs02}. 
The number of LO settings covering a given frequency is labeled as the redundancy.   For line-rich sources, based upon simulations, we estimate that a redundancy of $>$ 4 provides the needed fidelity for deconvolution.  We applied the standard
HIFI deconvolution using the {\it doDeconvolution} task within HIPE.
In Fig.~\ref{fig:dsb} we illustrate this method by presenting a Band 1a spectrum obtained with 3 (out of 4) of the LO settings within a reduced frequency range and the final SSB spectrum.

\begin{figure}
\resizebox{\hsize}{!}{\includegraphics{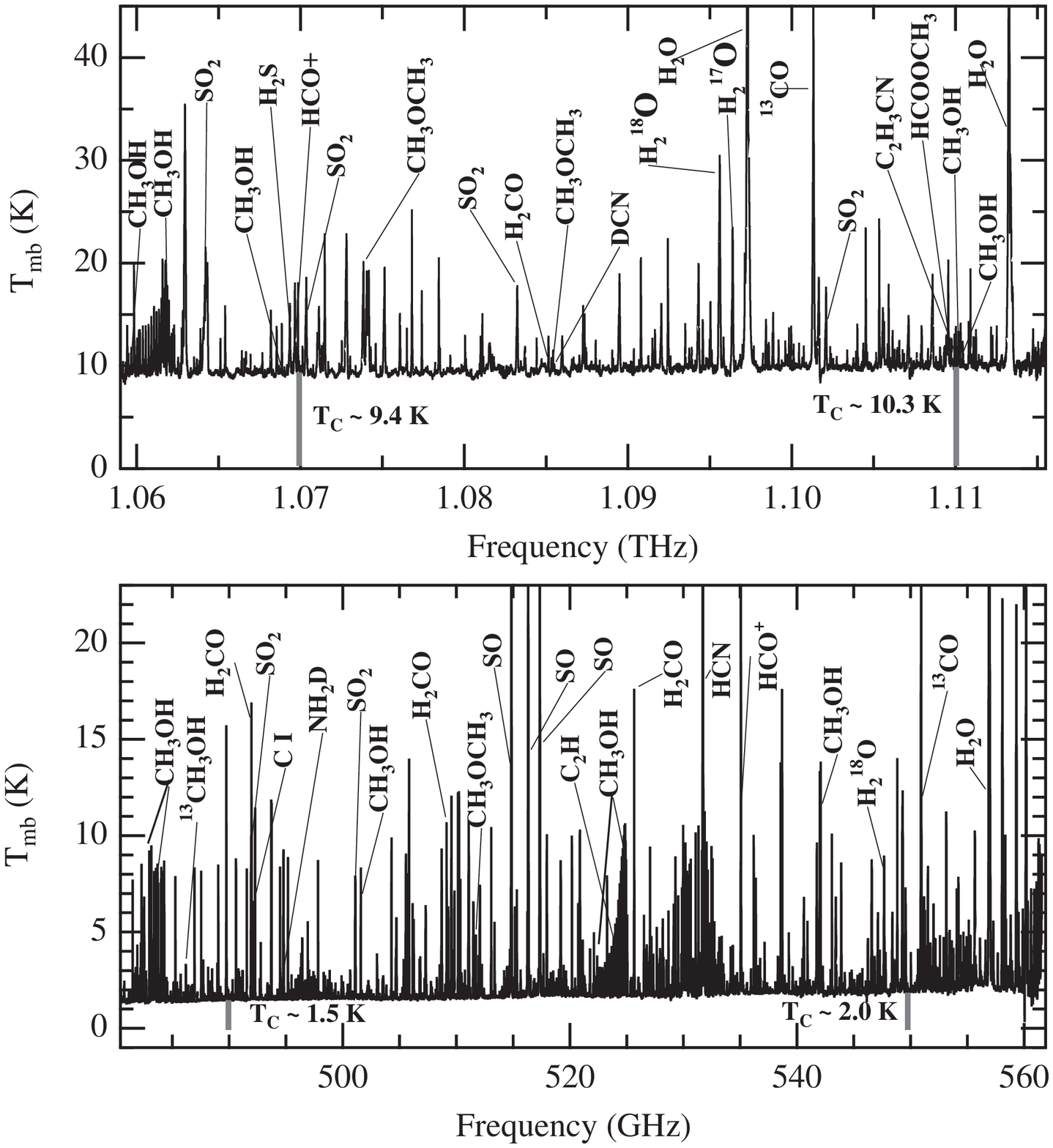}}
\caption{{\em Top:} HIFI spectral scan of Orion KL in Band 4b. {\em Bottom:} HIFI spectral scan of Orion KL in Band 1a.   Strong lines in both spectra are identified.  At the bottom of each panel we give the continuum level seen in the data at the ends of the bands.  For line identification we made use of the myXCLASS program
  (http://www.astro.uni-koeln.de/projects/schilke/XCLASS) which accesses
  the CDMS \citep[][http://www.cdms.de]{muller01, muller05} and JPL
  \citep[][http://spec.jpl.nasa.gov]{pickett98} molecular data bases for line assignment.
}
\label{fig:spec}
\end{figure}

We present here the single sideband data of Band 1a (480 -- 560 GHz)
and 4b (1.06 -- 1.115 THz) obtained as part of the {\em Herschel} science demonstration phase; however we use data from the other bands in our discussion of HD$^{18}$O.   The survey spectra are shown in Fig.~\ref{fig:spec} and have an angular resolution of $\theta_{1a} \sim 40''$ and $\theta_{4b} \sim 20''$ .  A blow up of a region of the spectrum in 1a is shown in Fig.~\ref{fig:dsb}, which illustrates the fidelity of the data.   The rms in the center of band 1a is 20 mK and 67 mK in 4b,  obtained using a velocity resolution of 1 km s$^{-1}$  and $\eta_{mb} = 0.7$.

   %
   
%
 %

\section{Results}

The spectral scans present a fantastic {\em Herschel} legacy but also a daunting data product with tens of thousands of lines seen above the noise.     Here we isolate two areas for initial focus: the line-to-continuum ratio in Orion KL as a function of frequency and the D/H ratio of water in Orion KL.

\subsection{Dust emission and the line-to-continuum ratio}

\begin{figure}
\resizebox{\hsize}{!}{\includegraphics{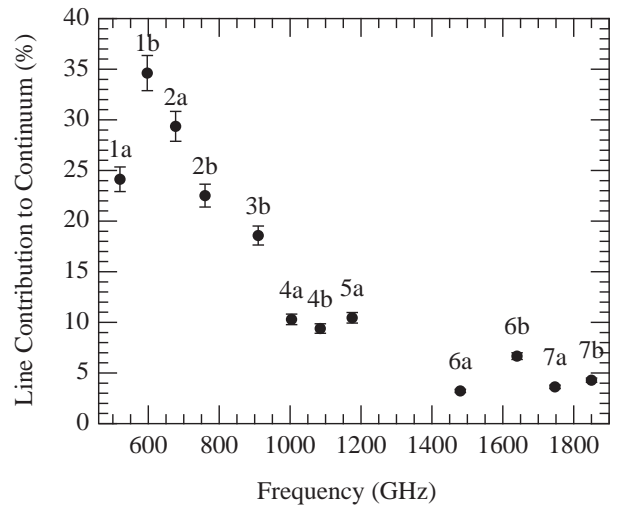}}
\caption{Percentage of the continuum flux contributed by line emission (line $+$ continuum/continuum) in each of the observed HIFI bands (1a, 1b, 2a, 2b, 3b, 4a, 4b, 5a, 6a, 6b, 7a, 7b) towards Orion KL.  We assume a 5\% relative error on the determination which is show as the error bars.}
\label{fig:lc}
\end{figure}

The HIFI spectral scans obtained with DBS mode detect both continuum and line emission.  In the Orion (and Sgr B2) data, we see evidence of a rise in the continuum within a given sub-band (see Fig.~\ref{fig:spec}).    The continuum is detected even in the slow (0.125 Hz) chop setting  used for bands 1 -- 5.
In Fig.~\ref{fig:spec} the continuum is $\sim$1.7~K at 520 GHz and $\sim$10~K at 1.1 THz.  This represents a measurement of the continuum flux towards Orion without line contamination; however the telescope beam, and hence beam-source coupling, is changing with frequency, which complicates direct continuum analysis.  \citet{sbmp84} and \citet{groesbeck95} estimate that the contamination due to line flux at 230 GHz and 330 GHz is roughly $\sim 50$\%.    In Fig.~\ref{fig:lc} we provide an estimate of the line contribution to the continuum derived from the available HIFI data.   These values were estimated by averaging the flux within a given band, including the line contribution and the continuum rise and by dividing by the continuum strength in the band center.    If Orion KL is representative of other more distant massive star-forming cores, then lines may contribute up to $\sim 20-40$\% of the continuum below 1 THz.   Above 1 THz the line contamination sharply decreases owing to rising continuum and decreasing line emission \citep[see][]{crockett10}.

Given this estimate of the line contribution, we can test how well HIFI measures the dust continuum in DBS mode by comparison to ground-based measurements.
 Assuming an aperture efficiency of 0.70, we estimate a flux of  $\sim$ 680 Jy  at 550 GHz and 4100 Jy at 1.1 THz.
 The measured total flux (including lines) in bands 1a and 4b is 840 Jy and 4350 Jy, respectively.    The flux comparison is more reliable at 520 GHz  where we can convolve the ground-based data to a comparable beam size
  and extrapolate over a narrower range in frequency, assuming that the flux scales as $\nu^{3.5}$ \citep{lis98, dicker09}.    Using the 350 $\mu$m SHARC map of OMC-1 \citep{lis98},  convolved to a {\em Herschel} beam of 36$''$ we estimate the flux towards Orion~KL to be 5250 Jy.   Thus the flux at 577 $\mu$m (520 GHz) is $\sim$ 912 Jy.  This is comparable to the HIFI measurement and within the errors of the flux scaling.    Another consistency check regarding the HIFI continuum measurement is found in the detection of optically thick water lines towards Sgr~B2 \citep[see ][]{lis_op}.
  
\subsection{Detection of HD$^{18}$O in the Orion Hot Core}

It is well known that hot cores have enhanced levels of HDO/H$_2$O \citep{jacq90, gmw96}.   Deuterium fractionation is inefficient at the high temperatures (T $\sim 100 - 200$ K) characteristic of the hot cores \citep{millar_dfrac}, and these enhancements are believed to be fossil remnants imprinted in ices during earlier colder phases.   In the case of Orion KL and other hot cores, the water column is estimated from the $3_{13} - 2_{20}$ transition of H$_2^{18}$O and various HDO transitions, which can be observed from the ground \citep{tgp_h218o, jacq88, jacq90}.  The interpretation of H$_2^{18}$O emission is somewhat complicated by the lack of other transitions and by the fact that the $3_{13} - 2_{20}$ transition is blended with an SO$_2$ line in galactic hot cores \citep[in sources with smaller line widths, the lines can be separated, e.g.][]{vdt06}.   In the case of HDO, the low-energy  lines in Orion KL seen in ground based spectra are quite strong ($T_{mb} \sim 40 - 60$ K) when correcting for the source size of $\sim 10''$ \citep{jacq90}.  Given temperatures of 100 - 200 K, it is possible, or even likely, that these lines are optically thick.  In fact \citet{pardo01} detected the $2_{12} - 1_{1,1}$ and $1_{11} - 0_{00}$ transitions and found that the hot core emission in these lines is obscured by emission from the other spatial/velocity components seen towards this line of sight \citep[e.g. the outflow or plateau,][]{blake87, persson07}.  
 
 \begin{figure}
\resizebox{8cm}{!}{\includegraphics[angle=-90]{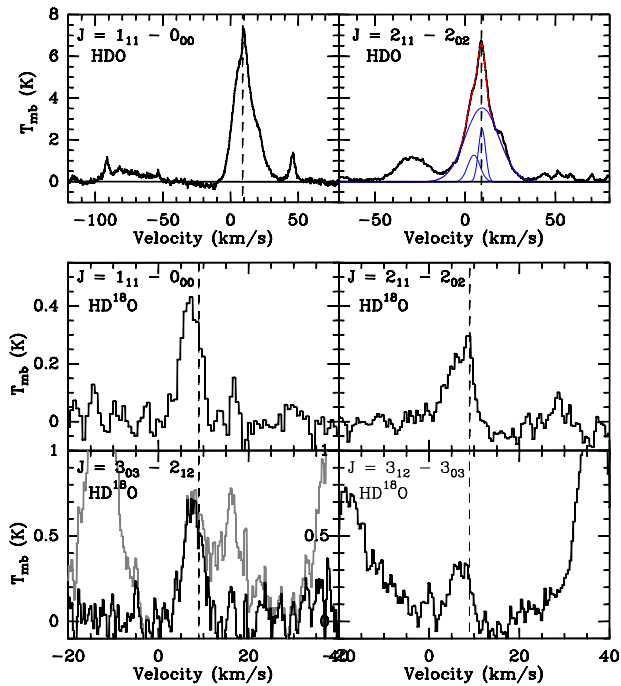}}
\caption{{Top 2 panels:} Selected transitions of HDO detected in HEXOS scans of Orion KL.  The panel for HDO $2_{11} - 2_{02}$ shows the typical fit with 3 spatial velocity components.  The narrow component is centered on 9 km s$^{-1}$ and is associated with the compact ridge, the very broad component is the plateau, and the hot core is at 5--6 km s$^{-1}$.   {Lower 4 panels:} Spectra of 4 transitions of HD$^{18}$O.  All data are shown on the T$_{mb}$ scale.  The $3_{03} - 2_{12}$ transition of HD$^{18}$O is blended with another line.  We show the original spectrum in grey and the unblended isolation of the HD$^{18}$O emission as the dark line.}
\label{fig:hdo}
\end{figure}

  Unbiased spectral scans offer the opportunity for unexpected discoveries.  One such example is the detection of multiple transitions of weak HD$^{18}$O emission in the Orion hot core.    A sample of the detected transitions is shown in Fig.~\ref{fig:hdo}, along with two transitions of HDO.   The $2_{12} - 2_{02}$ transition of HDO shows the distinctive shape of HDO lines in Orion with at least three spatial/velocity components visible in the spectrum \citep[see also][this volume]{melnick10}.    The Orion KL spectrum is at or near the line confusion limit and, as such, it is a question of whether these weak lines can be reliably assigned as HD$^{18}$O.   As can be seen in  Fig.~\ref{fig:hdo}, and listed in Table~1, the lines all are centered near 7-8 km/s with a similar width.  In addition, it is likely that the HDO emission is optically thick.  This is borne out in the ground state HDO $1_{11}-0_{00}$ spectrum, which has a strongly absorbed blue wing that hints at absorption below the continuum.  This is reminiscent of the spectrum of optically thick ground-state H$_2^{18}$O lines detected by {\em Herschel} \citep{melnick10}.   Finally, excitation models show that the detected transitions of  HD$^{18}$O are the strongest lines (that are not blended with other transitions in our data).  Thus we tentatively assign these features to HD$^{18}$O in Orion KL.

The detected lines are relatively broad ($\Delta v \sim 5-6$ km s$^{-1}$)  with an observed line center velocity of $\sim 7$ km s$^{-1}$, which is in between the expected velocity of the compact ridge of 9 km s$^{-1}$ and hot core of 5--6 km s$^{-1}$ \citep[see ][]{persson07}.   Similarly they are too narrow for the outflow component.  To examine this question we used HDO emission spectra (see top panel in Fig.~\ref{fig:hdo}) to fix the expected line parameters for HD$^{18}$O and explore which component dominates the emission.  In this fashion, we estimate that typically 60--80\% of the HD$^{18}$O emission arises in the hot core, and assign the detections to that component.   The remainder ($\sim 20-40$\%) could be attributed to the compact ridge, or blends from other interfering lines for some transitions.
  
  \citet{melnick10} analyze the numerous H$_2$O and H$_2^{18}$O lines detected by {\em Herschel}/HIFI in Orion KL.   They derive a total H$_2^{18}$O column of N(o+p H$_2^{18}$O) $= 0.5 - 2 \times 10^{16}$ cm$^{-2}$ with a source size of 8$''$.  This assumes n(H$_2$) = 1 $\times 10^7$ cm$^{-3}$ and T = 150~K.  In addition, the water emission has a significant contribution from radiative excitation.   An LTE analysis of the HD$^{18}$O emission with these parameters gives a total column of N(HD$^{18}$O) $= 1.6 \times 10^{14}$ cm$^{-2}$.    We consider this uncertain, since we assume LTE, do not account for radiative excitation, and attribute the emission solely to the hot core.
 Thus the D/H ratio of water in the hot core is $\sim $0.02.  This limit is significantly higher than estimated by \citet{jacq90} and
 \citet{persson07} (HDO/H$_2$O $\sim 10^{-3}$); however, it is close to the ratio estimated by \citet{persson07} for the compact ridge ($\sim 3 \times 10^{-2}$).    A more definitive analysis will be performed  using all the HDO and water lines.  Regardless, the possible detection of HD$^{18}$O hints that the D/H ratio of water in hot cores should be closely examined.

 \begin{table}
      \caption[]{HD$^{18}$O  Line Parameters}
\centering
\label{tab1}
\begin{tabular}{lllll}
\hline\hline
\multicolumn{1}{c}{$\nu$(GHz)} & 
\multicolumn{1}{c}{Transition} & 
\multicolumn{1}{c}{$\int T_{mb}d{\rm v}$} &
\multicolumn{1}{c}{v} & 
\multicolumn{1}{c}{$\Delta$v} \\
&& (K km s$^{-1}$) & (km s$^{-1}$) & (km s$^{-1}$) \\\hline
592.402 & $2_{11}-2_{02}$ & 1.7 (0.1) & 7.1 (0.1) & 6.4 (1.2) \\
746.476 & $3_{12}-3_{03}$ & 2.9 (0.2) & 6.7 (0.2) & 6.3 (0.3) \\
883.189 & $1_{11}-0_{00}$ & 2.1 (0.1) & 7.1 (0.1) & 4.9 (0.3) \\
994.347\tablefootmark{a} & $3_{03}-2_{12}$ & 3.2 (0.2) & 7.9 (0.2) & 4.7 (0.3) \\\hline
\end{tabular}
\tablefoot{Two other transitions ($2_{11}-1_{10}$, $2_{02}-1_{11}$) might be present, but are not listed due to significant line blending. 
\tablefoottext{a}{Blended with unidentified line; emission clearly isolated using Gaussian fits (e.g. Fig.~\ref{fig:hdo}).}
}
\end{table}

\begin{acknowledgements}
  HIFI has been designed and built by a consortium of institutes and university departments from across 
Europe, Canada, and the United States under the leadership of SRON Netherlands Institute for Space
Research, Groningen, The Netherlands, and with major contributions from Germany, France, and the US. 
Consortium members are: Canada: CSA, U.Waterloo; France: CESR, LAB, LERMA,  IRAM; Germany: 
KOSMA, MPIfR, MPS; Ireland, NUI Maynooth; Italy: ASI, IFSI-INAF, Osservatorio Astrofisico di Arcetri- 
INAF; Netherlands: SRON, TUD; Poland: CAMK, CBK; Spain: Observatorio Astron—mico Nacional (IGN), 
Centro de Astrobiolog'a (CSIC-INTA). Sweden:  Chalmers University of Technology - MC2, RSS \& GARD; 
Onsala Space Observatory; Swedish National Space Board, Stockholm University - Stockholm Observatory; 
Switzerland: ETH Zurich, FHNW; USA: Caltech, JPL, NHSC.   The HEXOS team also is grateful to the HIFI instrument team for building a fantastic instrument.
Support for this work was provided by NASA through an award issued by JPL/Caltech.
\end{acknowledgements}

\bibliography{/Nirgal1/ebergin/tex/bib/ted}

\Online 
\appendix

\renewcommand{\thefigure}{\arabic{figure}}
\setcounter{figure}{1}

\begin{figure*}
\centering
\resizebox{\hsize}{!}{\includegraphics{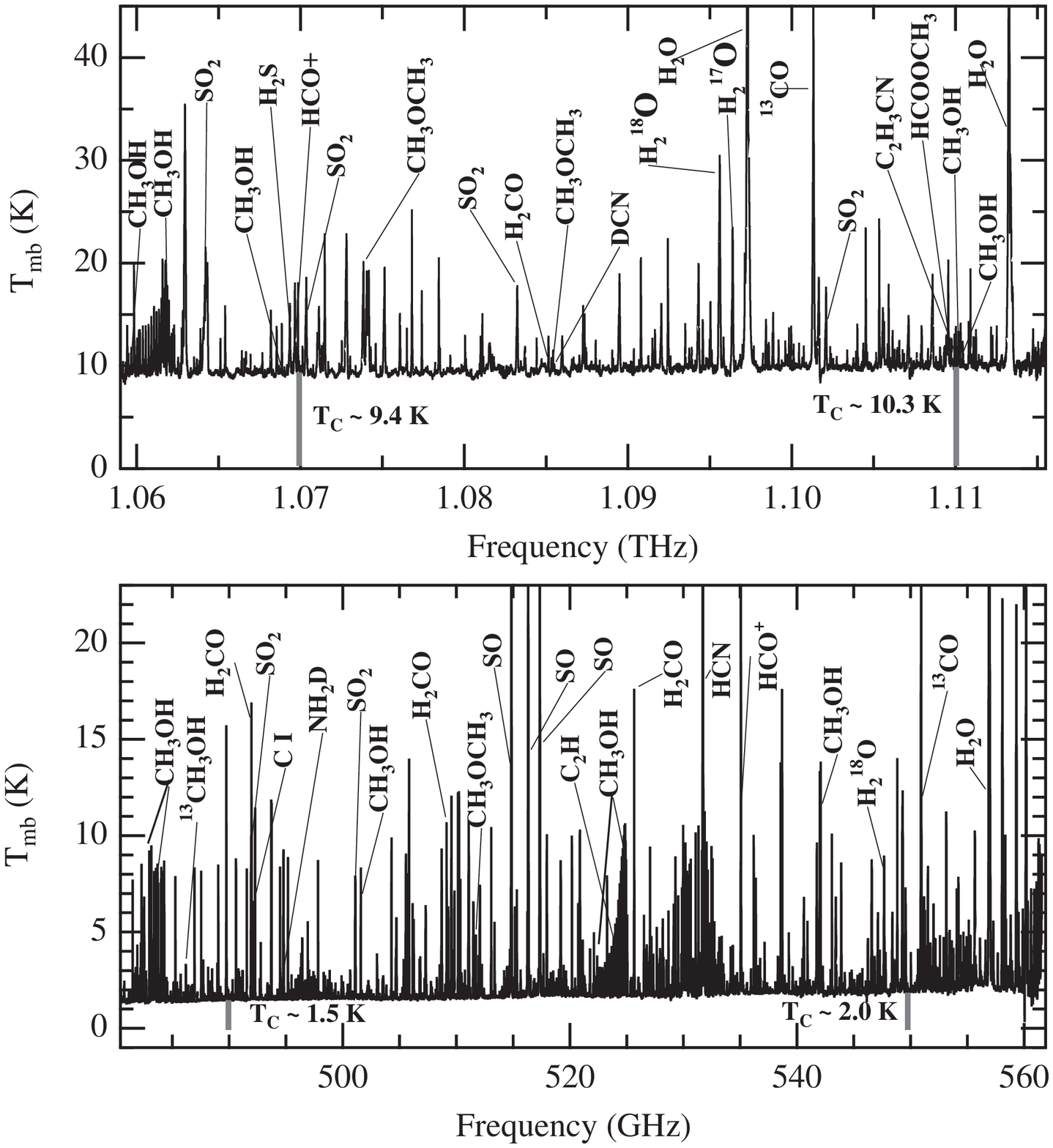}}
\caption{Larger version of Fig.~\ref{fig:spec}.  {\em Top:} HIFI spectral scan of Orion KL in Band 4b. {\em Bottom:} HIFI spectral scan of Orion KL in Band 1a.   Strong lines in both spectra are identified.  At the bottom of each panel we give the continuum level seen in the data at the ends of the bands.  We made use of the myXCLASS program
  (http://www.astro.uni-koeln.de/projects/schilke/XCLASS) which accesses
  the CDMS \citep[][http://www.cdms.de]{muller01, muller05} and JPL
  \citep[][http://spec.jpl.nasa.gov]{pickett98} molecular data bases for line assignment.
}
\label{appfig}
\end{figure*}

\end{document}